# Systematic Derivation of Reliable Wake Functions for Complex Structures from Mesh-Based Wakefield Simulations

Chih-Kai Liu[1], Wai-Keung Lau[2], Shih-Hung Chen[1*], Wei-Yuan Chiang[2†]

[1]Department of Physics, National Central University, Taoyuan 320317, Taiwan

[2]National Synchrotron Radiation Research Center, Hsinchu 300092, Taiwan

[*] chensh@ncu.edu.tw ; [†] chiang.wy@nsrrc.org.tw

## Abstract

Wakefield calculations are essential for analyzing beam-driven electromagnetic structures in accelerators. Although analytical wake functions are available for simple symmetric structures, complex geometries generally require mesh-based electromagnetic simulations, which provide finite-bunch wake potentials rather than point-charge wake functions directly. In this study, we present a systematic deconvolution-based method for extracting reliable wake functions from numerically calculated wake potentials using the prescribed drive-bunch distribution. The method is validated with a rectangular dielectric-lined waveguide (DLW), where the extracted longitudinal wake functions agree well with analytical solutions in both the short- and long-range regimes when the drive bunch is sufficiently short. The extracted wake function is further implemented in particle-tracking simulations, producing phase-space distributions consistent with those obtained using a built-in analytical wake-function model.

The method is also applied to a modified rectangular DLW with a non-uniform horizontal dielectric distribution. The extracted longitudinal and transverse wake functions and corresponding beam impedances show that the dominant deflecting wakefield can be substantially reduced without significantly degrading the longitudinal wakefield. These results demonstrate the reliability and applicability of the proposed method for complex dielectric-loaded structures.

# I. INTRODUCTION

When a relativistic charged particle beam traverses through a conducting enclosure that may contain dielectric materials, it interacts with surrounding structure and generate electromagnetic wakefields. These wakefields with some specific spatial and temporal structures might disturb the dynamics of the trailing charged particles. Comprehensive study of the properties of wakefields excited in these dielectric-filled structures is motivated by studies of collective beam instabilities in accelerator systems [1-2]. Electromagnetic structures can also be deliberately designed to generate wakefields for accelerating trailing beams.

The concept of wakefield acceleration can be traced back to the 1940s [3-5]. Proof-of-principle experiments conducted in the 1950s demonstrated beam acceleration at the keV/m scale [6-7]. Following these fundamental works, researchers started to investigate the feasibility of wakefield accelerators with field gradients that are much higher than conventional microwave linacs [8-9]. Circular waveguides with periodic structures or dielectric liners are commonly used as wakefield-generating structures. Periodic structures are often optimized to match the electron beam size, thereby enhancing wakefield-driven acceleration [10]. In contrast, dielectric-lined waveguides (DLWs) rely on Čerenkov radiation excited by a relativistic electron bunch propagating through the dielectric medium. The radiation field is guided and reflected by the conducting boundary of the structure, allowing it to interact with trailing particles and accelerate them. DLW structures have been demonstrated to generate accelerating gradients as high as 100 MV/m [11], demonstrating their potential for the development of compact accelerator systems.

Recently, wakefields in electromagnetic structures have also been utilized as versatile tools for phase-space manipulation and electron-beam diagnostics in advanced accelerator systems [12–22]. In practice, manipulation and diagnostics of ultrashort, high-brightness electron beams are commonly performed using parallel-plate corrugated waveguides or DLW structures. Given the manufacturing challenges associated with meter-scale corrugated waveguides, the strong wakefields generated in DLWs enable the design of compact structures with relatively simple geometries. In practical applications, DLW dechirpers can be used to reduce correlated energy spread, with dechirping strengths as high as ~ 100 keV/μm. For slice energy-spread measurements, transverse wakefields can introduce a longitudinally dependent transverse kick, thereby converting the bunch's longitudinal structure into a measurable transverse displacement with head-to-tail separations of approximately 10 mm.

Wakefields can be used for beam acceleration, dechirping, and beam deflection, enabling the development of application-specific accelerator components. However, wakefields excited by the drive beam can also produce detrimental effects. In particular, transverse wakefields can act on trailing particles or on later portions of the same bunch, leading to beam breakup (BBU) instabilities. Therefore, accurate determination of the wake function of a given device is essential for the advance evaluation and mitigation of instabilities in accelerators and storage rings. The method presented in this work provides a systematic approach for achieving this goal.

The wake function is defined as the path integral of the electromagnetic force associated with wakefields excited by a unit point charge. In the ultra-relativistic limit, it is determined solely by the geometry and material properties of the wakefield-generating structure. Analytical wake functions, however, are available only for a limited number of simple geometries. For complex structures, the wake function must generally be reconstructed from numerically computed wake potentials through deconvolution. This is because the wake function corresponds to the response of a unit point charge, which is mathematically represented by a delta-function excitation and therefore cannot be directly resolved in mesh-based simulations. As pointed out by Zotter and Kheifets [23], the drive-bunch length used in such simulations must be several times larger than the mesh size. Consequently, mesh-based simulations provide wake potentials associated with finite drive-bunch distributions rather than the wake function itself. The reconstructed wake function can then be used as input to well-developed multi-particle tracking codes, such as ELEGANT [24] and IMPACT [25], to evaluate its impact on beam dynamics.

Wake potentials can be computed using electromagnetic simulation codes such as CST [26], T3P [27], ECHO [28], PBCI [29], and GdfidL [30]. Accurate numerical evaluation often requires fine spatial resolution, appropriate integration treatment, and sufficiently long computational domains, particularly when both short- and long-range wake information is required. In practical wakefield problems, meter-scale structures may be driven by electron beams of micrometer-scale bunch lengths, so resolving the beam and the associated wakefields in full electromagnetic simulations can lead to prohibitively large numbers of computational cells and substantial computational cost. Therefore, the choice of simulation code is essential. CST Studio Suite is a parallelized software package that supports direct CAD-file import and provides a user-friendly interface. Although CST does not employ a moving-window technique, unlike T3P and GdfidL, this limitation is not necessarily a disadvantage for calculations requiring long-range wake information. In addition, CST provides a three-dimensional particle-in-cell simulation tool for studying the interaction between structure-generated electromagnetic fields and charged particle beams. For example, CST [31] and VSim [32–33] have been used to study energy dechirpers for reducing correlated energy spread in electron beams, although such simulations require substantial computational resources.

In this work, we propose a systematic method for extracting accurate wake functions from wake potentials numerically calculated using CST for electromagnetic structures driven by Gaussian bunches. Dielectric-lined waveguide (DLW) beam dechirpers are employed as representative examples to validate the proposed approach. Specifically, the wake functions extracted from CST-calculated wake potentials are compared with analytical wake functions for rectangular DLW structures. The results show good agreement in both the short- and long-range regimes. The extracted longitudinal wake function is further implemented in IMPACT-T [34] multi-particle tracking simulations, and the resulting phase-space distributions are compared with those obtained using the built-in analytical wake function. This comparison further confirms the applicability of the proposed method to beam-dynamics studies.

Previous studies on rectangular DLW structures have shown that even a small transverse beam offset can induce substantial transverse wakefields. Theoretical analyses have indicated that, in addition to the longitudinal wakefield, rectangular DLWs can generate transverse wakefields that deflect the beam away from the longitudinal axis [35]. These transverse wakefields may significantly affect trailing bunches and lead to transverse beam instabilities in accelerators. Earlier mitigation strategies relied on two orthogonally oriented planar DLW structures to compensate the transverse impedance; however, this approach increases the overall device length and introduces additional challenges in alignment and mechanical stabilization. Other studies have suggested that transverse wakefields can be reduced using flat drive beams [36, 37], but this imposes additional constraints on beamline design.

With the methodology developed in this work, complex geometrical and material parameters can be systematically optimized to mitigate transverse wakefields without significantly degrading the longitudinal wakefield. Based on this capability, we propose a novel DLW design in which a discontinuous dielectric distribution is introduced in the transverse direction to substantially reduce the transverse wakefield.

The remainder of this paper is organized as follows. Section 2 briefly reviews the concept of wakefields and the definitions of wake function and wake potential. It also summarizes the analytical wake functions for rectangular DLW structures, which were initially studied by Tremaine [36] and later fully developed by Mihalcea [37]. These analytical wake functions are used as references for comparison with wake functions extracted from CST-calculated wake potentials for Gaussian drive bunches with various bunch lengths. Section 3 presents the procedure for extracting wake functions from numerically calculated wake potentials. Section 4 verifies the applicability of the proposed methodology to multi-particle beam-dynamics simulations and discusses its application to transverse wakefield mitigation. Section 5 presents the conclusions.

## II. ANALYTICAL WAKEFIELD FORMULATION FOR RECTANGULAR DLW STRUCTURES

Dielectric-lined waveguides (DLWs) are metallic waveguides partially loaded with dielectric materials and are widely used in accelerator systems for beam manipulation and wakefield-driven acceleration. When a relativistic charged particle passes through a DLW, coherent Čerenkov radiation is excited in the dielectric structure and generates wakefields that act on trailing particles. As discussed in the previous section, a rectangular DLW structure is used here as an example to demonstrate the validity of the proposed method for extracting wake functions from mesh-based simulations. For this purpose, we first review the basic definitions of wakefield theory and the analytical wakefield solutions for rectangular DLW structures.

The longitudinal wakefield is of particular importance because it determines the energy modulation experienced by witness charged particles or by different longitudinal slices of the same bunch. The wake potential is obtained from the longitudinal wake function. Consider a source particle with charge $q'$ moving through the structure and a trailing test particle with charge $q$ located at a longitudinal distance $s$ behind the source particle. The energy change of the test particle can be written as

$$\Delta U = qq'W_{\parallel}(\mathbf{r}_{\perp}, s), \tag{1}$$

where $W_{\parallel}(\mathbf{r}_{\perp}, s)$ is the longitudinal point-charge wake function, defined as the longitudinal wake potential per unit source charge, $\mathbf{r}_{\perp}$ denotes the transverse position of the test particle, while $s > 0$ specifies its longitudinal separation from the source particle. The corresponding point-charge wake function is obtained by integrating the longitudinal electric field induced by the source charge:

$$W_{\parallel}(\mathbf{r}_{\perp}, s) = \frac{1}{q'} \int_{-\infty}^{\infty} E_z\left(\mathbf{r}_{\perp}, z, \frac{s+z}{v}\right) dz, \tag{2}$$

where $E_z$ is the longitudinal electric field, $z$ is the coordinate along the beam-propagation direction, and $v$ is the particle velocity. In the ultra-relativistic limit, the wake function is primarily determined by the geometry and material properties of the DLW.

For a realistic finite bunch, the source charge is described by a longitudinal line-charge distribution $\lambda(s)$. For a Gaussian bunch, the normalized longitudinal distribution is given by

$$\lambda(s) = \frac{1}{\sqrt{2\pi}\sigma} \exp(-\frac{s^2}{2\sigma^2}), \tag{3}$$

where $\sigma$ is the rms bunch length, and $\lambda(s)$ satisfies

$$\int_{-\infty}^{\infty} \lambda(s) ds = 1. \tag{4}$$

The corresponding wake potential is obtained by convolving the point-charge wake function with the normalized longitudinal line-charge distribution:

$$W_{\lambda}(s) = -\int_{0}^{\infty} W_{\parallel}(s')\lambda(s-s')ds', \tag{5}$$

where $W_{\lambda}(s)$ denotes the longitudinal wake potential of the finite bunch. Equation (5) shows that the wake potential represents a smoothed form of the point-charge wake function. Therefore, when the wake potential and the bunch profile are known, the point-charge wake function can, in principle, be recovered through deconvolution [38,39], namely,

$$W_{\parallel}(s) = \mathcal{F}^{-1}\left\{\frac{\mathcal{F}[W_{\lambda}(s)]}{\mathcal{F}[\lambda(s)]}\right\}, \tag{6}$$

where $\mathcal{F}$ and $\mathcal{F}^{-1}$ denote the Fourier transform and inverse Fourier transform, respectively. In practice, this deconvolution procedure is sensitive to numerical noise, particularly for large longitudinal wavenumbers where $\mathcal{F}[\lambda(s)]$ becomes small. Therefore, accurate wake-potential calculations are essential for reliable wake-function extraction.

To characterize the deflecting force exerted on a trailing particle by the transverse wakefields excited by an off-axis drive particle, the dipole transverse wake function is defined as the integrated transverse Lorentz force normalized by the source charge and its transverse offset:

$$W_{\perp}(\mathbf{r}_{\perp}, s) = \frac{1}{q'd}\int_{-\infty}^{\infty} dz \left(\vec{E} + \vec{v} \times \vec{B}\right)_{\perp}, \tag{7}$$

where $d$ denotes the transverse displacement of the drive particle from the structure axis. To analyze the wakefield effects in the frequency domain, the corresponding coupling impedances are defined as the Fourier transforms of the wake functions. Longitudinal coupling impedance is thus defined as

$$Z_{\parallel}(\omega) = \int_{-\infty}^{\infty} d\tau W_{\parallel}(s) e^{-j\omega\tau}, \tag{8}$$

with $\tau = s/c$. And the corresponding transverse coupling impedance is defined as

$$Z_{\perp}(\omega) = \int_{-\infty}^{\infty} d\tau W_{\perp}(s) e^{-j\omega\tau}. \tag{9}$$

For simple DLW geometries, analytical wakefield solutions can be obtained by solving the corresponding waveguide eigenvalue problem subject to appropriate electromagnetic boundary conditions. These analytical solutions provide useful benchmark cases for validating numerically computed wake potentials.

In a rectangular DLW, as shown in Fig. 1, purely transverse electric (TE) or transverse magnetic (TM) modes generally do not exist. The eigenmodes are instead classified as longitudinal-section electric (LSE) modes and longitudinal-section magnetic (LSM) modes. At the dielectric-vacuum interface, LSE modes satisfy $E_y = 0$, whereas LSM modes satisfy $H_y = 0$. Assuming that the center of the rectangular DLW is located at $(x, y) = (0,0)$, as shown in Fig.

1(a), the longitudinal electric field of the $(m, n) - th$ longitudinal-section mode can be written in the same mathematical form for both LSE and LSM modes, following Mihalcea et al. [36]:

$$E_{z;m,n}^{LS,\zeta} = \begin{cases} E_{0;m,n}^{LS,\zeta} \, cos(\kappa_{x,m}x) \, cosh(\kappa_{x,m}y) & 0 < y < a \\ E_{0;m,n}^{LS,\zeta} \frac{cosh(\kappa_{x,m}a)}{sin[\kappa_{y,n}(b-a)]} cos(\kappa_{x,m}x) \, sin[\kappa_{y,n}(b-y)] & a < y < b \end{cases} . \tag{10}$$

Here, $E_{0;m,n}^{\mathrm{LS},\zeta}$ denotes the modal amplitude of the (m, n)-th longitudinal-section mode, where $\zeta = M$ corresponds to the LSM modes and $\zeta = E$ corresponds to the LSE modes. The parameters $a$ and $b$ denote the distances from the waveguide center to the inner and outer boundaries of the dielectric-lined region, respectively. The transverse wave numbers $\kappa_{x,m}$ and $\kappa_{y,n}$ in the $x$- and $y$-directions, respectively, of the $(m, n) - th$ longitudinal-section (LS) mode are determined by the corresponding dispersion relation:

$$\coth\,(\kappa_{x,m}a) \cot\,[\kappa_{y,n}(b-a)] = \frac{\kappa_{y,n}}{\epsilon_r \kappa_{x,m}} \tag{11}$$

for LSM modes and

$$\coth\,(\kappa_{x,m}a) \cot\,[\kappa_{y,n}(b-a)] = -\frac{\kappa_{x,m}}{\kappa_{y,n}} \tag{12}$$

for LSE modes. The analytical modal amplitudes of the LSM and LSE modes can be written as

$$E_{0;m,n}^{LS,M} = \frac{\cosh(\kappa_{x,m}y_0)}{2\epsilon_0} \left[ \frac{\sinh(2\kappa_{x,m}a)}{2\kappa_{x,m}} + \frac{\epsilon_r \cosh(\kappa_{x,m}a)}{sin^2[\kappa_{y,n}(b-a)]} \left\{ \frac{b-a}{2} \left(1 + \frac{\epsilon_r \kappa_{x,m}^2}{\kappa_{y,n}^2}\right) - \frac{\sin[2\kappa_{y,n}(b-a)]}{4\kappa_{y,n}} \left(1 - \frac{\epsilon_r \kappa_{x,m}^2}{\kappa_{y,n}^2}\right) \right\} \right]^{-1} \tag{13}$$

and

$$E_{0;m,n}^{LS,E} = \frac{\cosh(\kappa_{x,m}y_0)}{2\epsilon_0} \left[ \frac{\sinh(2\kappa_{x,m}a)}{2\kappa_{x,m}} + \frac{\cosh(\kappa_{x,m}a)}{sin^2[\kappa_{y,n}(b-a)]} \left\{ \frac{b-a}{2} \left(\epsilon_r + \frac{\kappa_{y,n}^2}{\kappa_{x,m}^2}\right) - \frac{\sin[2\kappa_{y,n}(b-a)]}{4\kappa_{y,n}} \left(\epsilon_r - \frac{\kappa_{y,n}^2}{\kappa_{x,m}^2}\right) \right\} \right]^{-1} \tag{14}$$

Here, $\epsilon_0$ is the vacuum permittivity, $\epsilon_r$ is the relative permittivity of the dielectric material, and $y_0$ is the transverse position of the driving particle.

The total longitudinal electric field in the rectangular DLW is obtained by superposing the contributions of all relevant LSM and LSE eigenmodes:

$$E_z(\mathbf{r}_\perp, z, t) = \sum_{m,n} \left[ E_{z;m,n}^{LS,M}(\mathbf{r}_\perp, z, t) + E_{z;m,n}^{LS,E}(\mathbf{r}_\perp, z, t) \right]. \tag{15}$$

The analytical wake function and wake potential can then be evaluated by substituting Eq. (15) into Eqs. (2) and (3). The rectangular DLW examples demonstrate that analytical wakefield calculations are possible when the geometry is sufficiently simple and the eigenmode expansion can be derived explicitly. The analytical solution can be used to validate wake potentials and wake function extracted through the CST-based procedure.

## III. NUMERICAL DECONVOLUTION PROCEDURE FOR WAKE-FUNCTION EXTRACTION

Practical accelerator structures often include three-dimensional geometries, finite-length effects, transitions, couplers, ports, multiple dielectric materials, and other non-ideal features. In such cases, approximate analytical models are generally insufficient to describe the resulting wakefields accurately; therefore, mesh-based electromagnetic simulations are required to calculate the wake potentials.

In this work, wake potentials are computed using the PIC Studio module of CST Studio Suite, which is based on the finite-difference time-domain (FDTD) method. Because a true point-charge excitation cannot be resolved on a finite mesh, the structure is driven by a finite Gaussian bunch with a prescribed longitudinal charge distribution. The simulation therefore yields the finite-bunch wake potential $W_\lambda(s)$, from which the corresponding point-charge wake function can be extracted through the deconvolution procedure described in Eq. (6).

The accuracy of the extracted wake function depends directly on the quality of the simulated wake potential. Therefore, the mesh resolution, bunch length, wakefield integration step, boundary treatment, and port design must be chosen carefully. In particular, since analytical solutions are usually derived for infinitely long waveguides, the simulated structure must be sufficiently long, and perfectly matched layer (PML) boundaries are applied at the longitudinal ends to suppress artificial reflections. Proper selection of these numerical settings reduces numerical dispersion, reflection artifacts, and high-frequency noise, thereby enabling reliable wake-function extraction.

### III-1. NUMERICAL DLW MODELS

CST studio suite provides a convenient platform for constructing complex three-dimensional geometries and importing CAD models generated by external design tools. It also supports the definition of materials with constant or spatially varying properties. These capabilities are particularly useful for modeling accelerator components with complex geometrical features and material distributions, such as dielectric-lined waveguides.

In this section, a rectangular dielectric-lined waveguide is used as a benchmark example to demonstrate the procedure for wake potential calculations in CST. Figure 1(a) shows the side view of the structure.

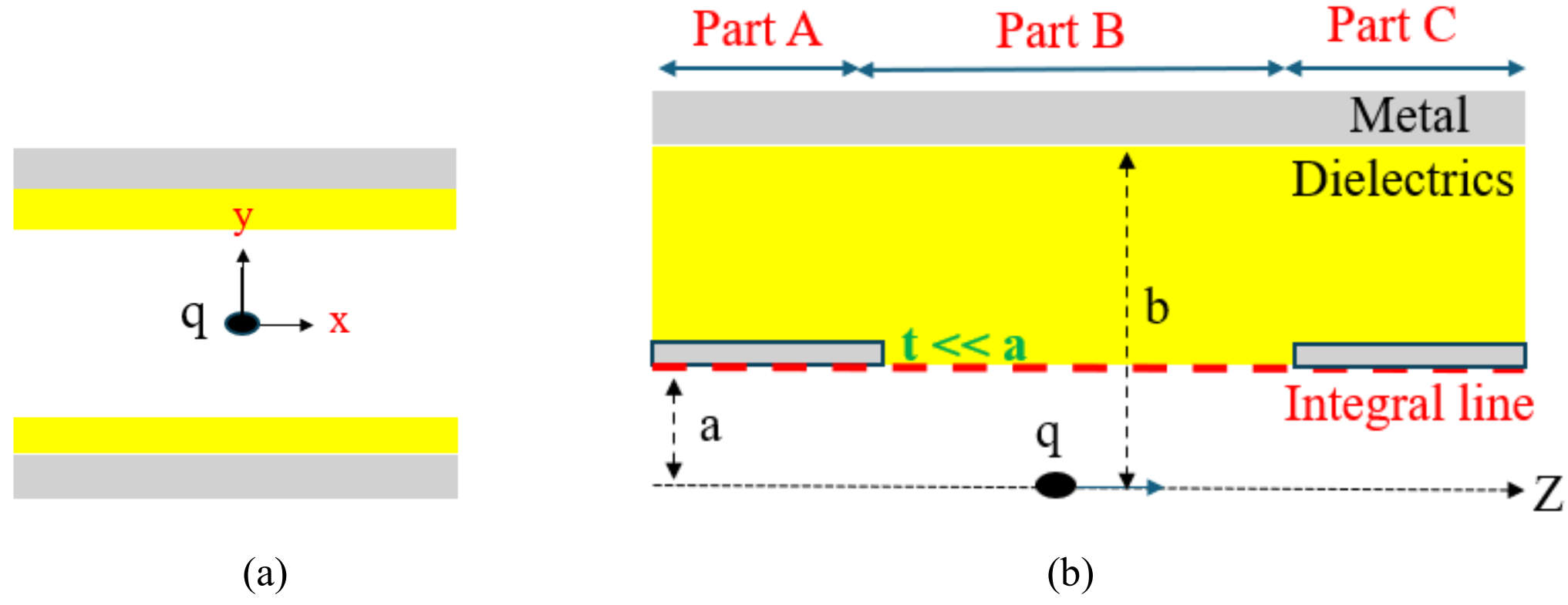

FIGURE 1. (a) Side view of the rectangular dielectric-lined waveguide. (b) Schematic of the rectangular dielectric-lined waveguide used for comparison with the analytical solution. Parts A and C are extended waveguide sections introduced to reduce field discontinuities and suppress boundary effects, whereas Part B is the straight waveguide section used for the wake-potential calculation.

When a particle bunch enters or exits a DLW, it encounters geometric discontinuities at the structural boundaries. In numerical simulations, these discontinuities can generate additional wakefields that are not included in the analytical model, which assumes an ideal uniform waveguide. Therefore, the simulation geometry must be designed to minimize artificial wakefield contributions from the entrance and exit regions.

Figure 1(b) shows the schematic layout of the simulation model. The central region, Part B, is a uniform rectangular dielectric-lined waveguide section, with a vacuum-channel half-aperture of 2.5 mm and a dielectric-layer thickness of 2.5 mm. This region is used as the main section for calculating the wake potential. Since this region corresponds to the geometry assumed in the analytical model, the calculated wake potential can be directly compared with the analytical result.

Parts A and C are extended waveguide sections introduced to reduce wakefields generated by entrance and exit discontinuities and to better approximate the infinite-waveguide assumption. This modular design allows different DLW configurations to be studied by replacing only Part B while keeping the entrance and exit sections unchanged. The two longitudinal ends of the simulation domain are treated with absorbing boundaries to suppress reflected fields.

### III-2. CONVERGENCE TESTS FOR DRIVE-BEAM PARAMETERS

Convergence tests were performed to determine the numerical parameters required for reliable wake-potential calculations. A Gaussian longitudinal distribution was used as the drive-bunch profile in CST. The particle velocity was set to the speed of light, consistent with the analytical model, and the total bunch charge was set to $10^{-7}$ C. The computed wake potential was normalized by the bunch charge during CST post-processing. Unless otherwise specified, the *rms* bunch length was set to 1 $mm$, whose influence on wake-function extraction is discussed in the next section.

The current excitation was defined using the analytical injection scheme to ensure a smooth representation of the moving bunch current. The bunch resolution was controlled by the CST parameter "lines per sigma," which defines the number of mesh lines used to resolve one *rms* bunch length. Based on the convergence test, this value was set to 30 in the present simulations. For the wake-potential evaluation, the indirect test-beam method was used instead of direct field integration, since the latter is more sensitive to numerical dispersion for relativistic beams. In CST, the indirect test-beam method calculates the wake potential by integrating electric fields along a path near the beam tube rather than directly along the beam axis. This approach helps reduce numerical errors associated with field singularities near the $z$-axis. In addition, scattered fields propagating in the beam tube can affect the accuracy of the wake-potential calculation, especially when the integration path or computational domain is not properly chosen [40, 41].

Figure 2 shows the effects of two key numerical parameters on the calculated wake potential. Here, the short-range wake refers to the wakefield within the electron bunch, whereas the long-range wake refers to the wakefield behind the bunch. First, the length of the uniform DLW section affects the contribution from entrance and exit discontinuities. As this length increases, the simulated wake potential approaches the analytical solution more closely. As shown in Fig. 2(a), a uniform section length of $300\ mm$ is sufficient to obtain a converged wake-potential result.

The second parameter is the frequency-limit setting. Shorter bunches excite higher-order modes in the DLW; therefore, the simulation bandwidth must be sufficiently large to include the relevant modal spectrum. In CST, the characteristic maximum frequency is determined from the bunch length, and the selected frequency limit should exceed this value to avoid filtering high-frequency wake components. For the $1\ mm$ bunch used here, the corresponding maximum frequency is approximately $177.3$ GHz. Figure 2(b) shows that the wake-potential result changes significantly near $200$ GHz and remains nearly unchanged for higher frequency limits. Accordingly, a frequency limit above $200$ GHz was used in the present simulations.

TABLE I. Dependence of the excitation frequency spectrum calculated by CST on the rms bunch length.

| **Bunch Length [mm]** | **Excitation frequency spectrum [GHz]** |
| --- | --- |
| 1 | 177.3 |
| 0.6 | 295.5 |
| 0.3 | 591 |
| 0.1 | 1773 |
| 0.05 | 3546 |

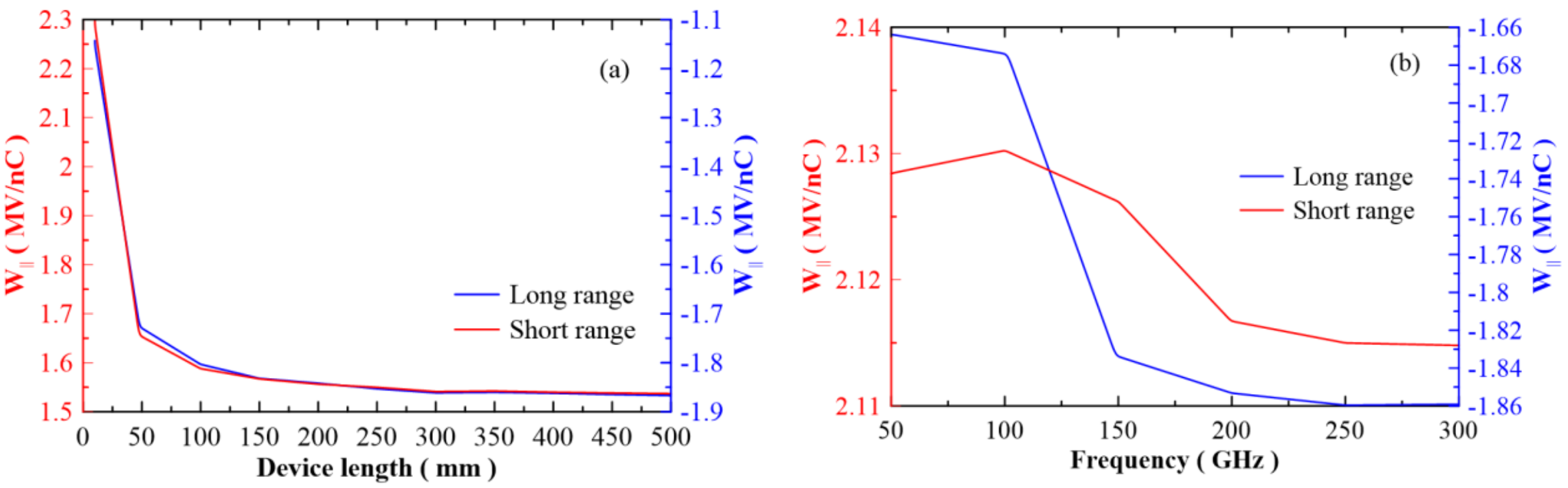

**FIGURE 2.** Convergence validation of the extracted short- and long-range wake functions with respect to numerical settings: (a) device length and (b) frequency-limit setting.

### III-3. WAKE-FUNCTION EXTRACTION FROM NUMERICAL WAKE POTENTIALS

According to the convolution relation in Eq. (4), the finite-bunch wake potential is obtained by convolving the point-charge wake function with the longitudinal bunch distribution in the longitudinal-coordinate domain. By applying the fast Fourier transform (FFT), this convolution is converted into multiplication in the frequency domain. The point-charge wake function can therefore be recovered by deconvolution, i.e., by dividing the frequency-domain wake potential by the frequency-domain bunch distribution:

$$W_{\parallel}(\omega) = \frac{W_{\lambda}(\omega)}{\lambda(\omega)}. \tag{16}$$

The corresponding wake function in the longitudinal-coordinate domain is then obtained by applying the inverse fast Fourier transform (IFFT) [38, 39].

In this work, the wake potential $W_{\lambda}(s)$ is first obtained from the CST wakefield simulation. The simulated wake potential and the prescribed Gaussian bunch distribution are then transformed into the frequency domain using the same FFT procedure in MATLAB. Zero padding is applied before the FFT procedure to improve the frequency resolution and reduce numerical artifacts. After performing the deconvolution in Eq. (15), the extracted wake function is converted back to the longitudinal-coordinate domain by applying the inverse fast Fourier transform (IFFT).

The complete procedure is summarized in Fig. 3. First, the key CST parameters are determined through convergence tests. The finite-bunch wake potential is then extracted from the CST simulation and deconvolved with the Gaussian bunch distribution to obtain the point-charge wake function. Finally, the extracted wake function is imported into IMPACT-T for multi-particle tracking simulations, and the resulting beam phase-space distribution is evaluated.

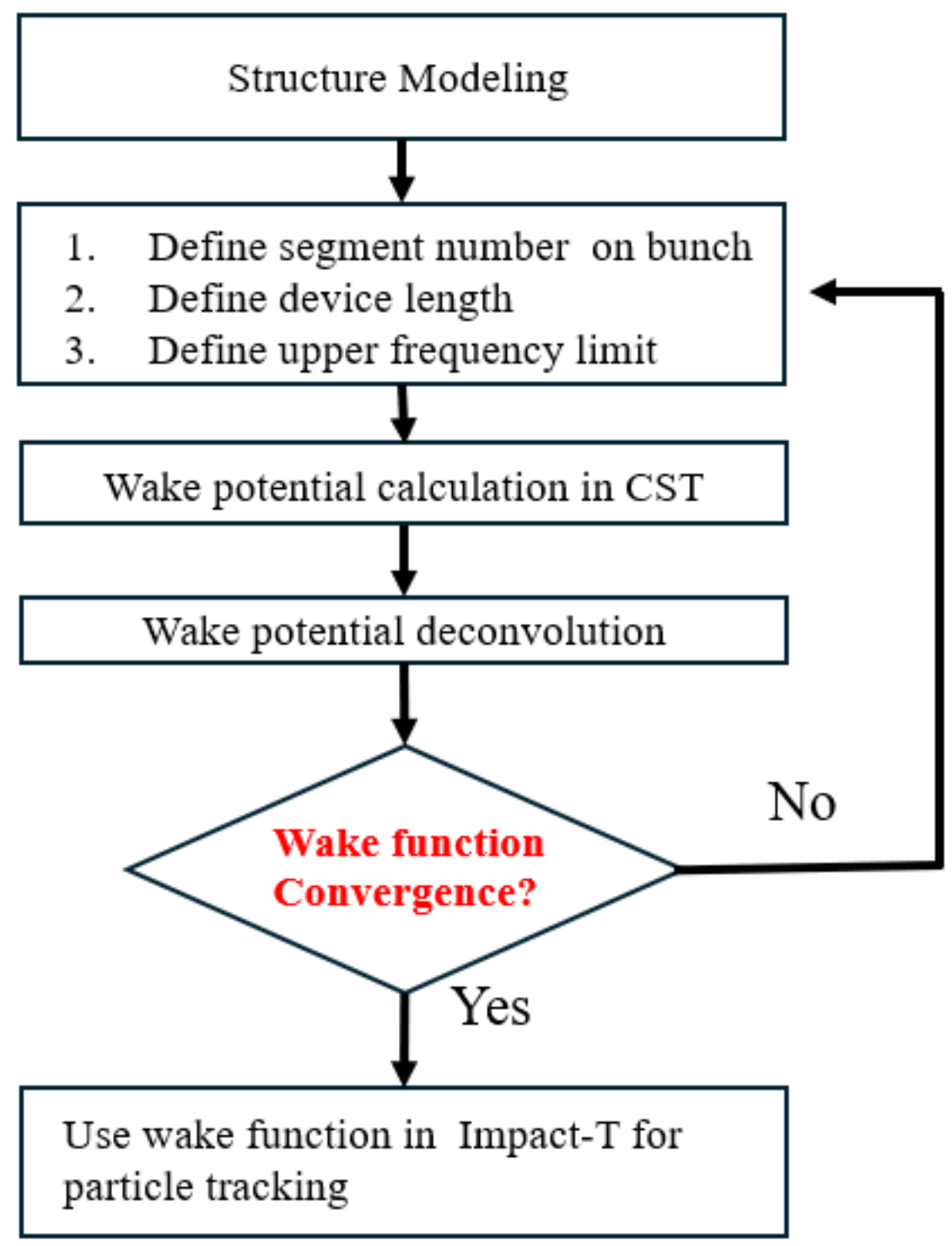

**FIGURE 3.** Flowchart of the CST-based convergence-test and wake-function extraction procedure.

## IV. MULTI-PARTICLE BEAM-DYNAMICS VALIDATION OF THE PROPOSED METHOD

Rectangular DLWs are used as benchmark cases to validate the proposed deconvolution method. The structural and material parameters of the rectangular DLW are shown in Fig. 4(a) and Table II, respectively. The vacuum half-gap is $a = 2.5\,mm$, and the outer boundary of the dielectric layer is located at $b = 5\,mm$, corresponding to a dielectric-layer thickness of 2.5 $mm$.

The total width of the rectangular waveguide is 10 $mm$, and the relative dielectric constant is $\varepsilon =$ 4. The *rms* bunch length used in the CST simulation is 1 $mm$.

Table II: Rectangular DLW parameter list

| Parameter | Values |
|---|---|
| a [mm] | 2.5 |
| b [mm] | 5 |
| w [mm] | 10 |
| Dielectric Constant | 4 |

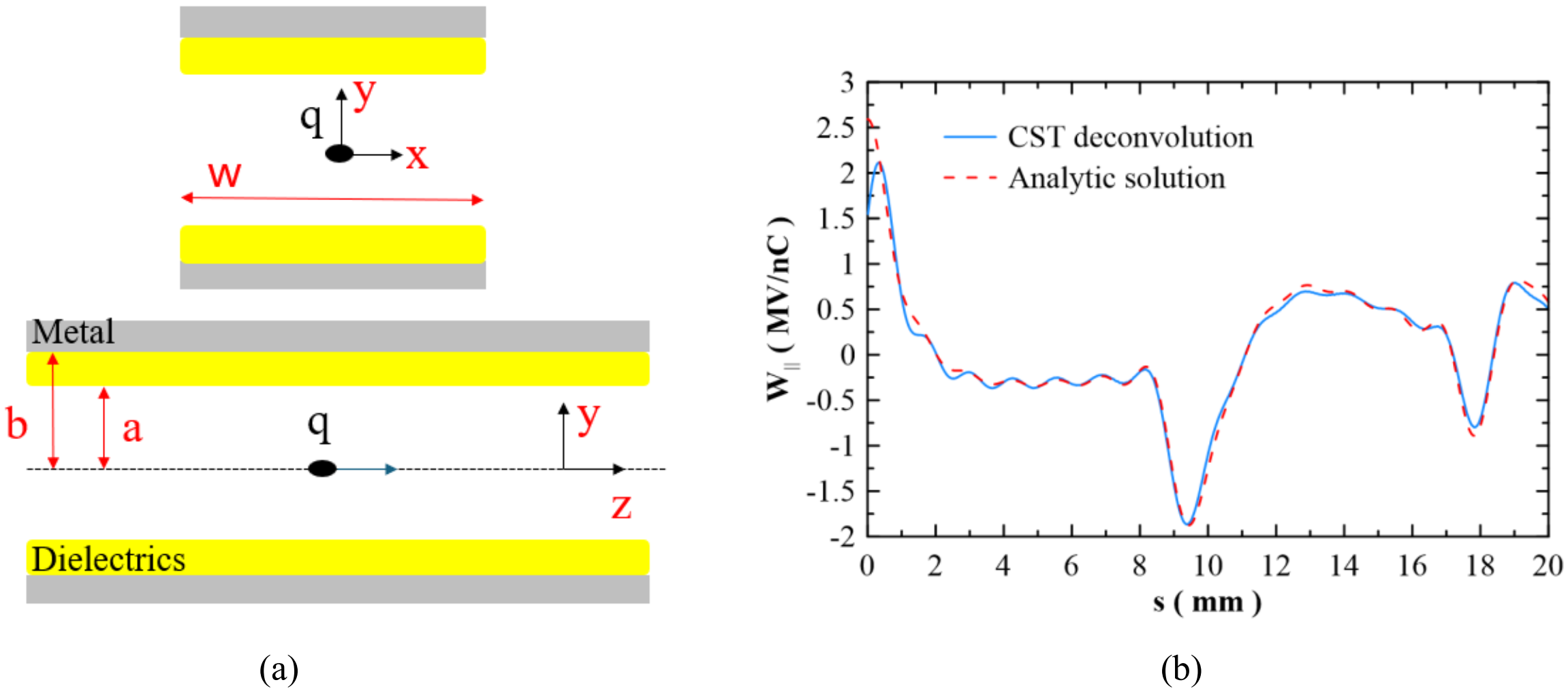

**FIGURE 4.** (a) Schematic layout of the flat dielectric-lined waveguide (DLW), where $w$ denotes the horizontal width and the dielectric layers are located in the vertical direction. The structural parameters are $a = 2.5\,mm$ , $b = 5\,mm$ , and $\varepsilon = 4$ . (b) Comparison between the CST-deconvolved wake function and the analytical wake function.

Figure 4(b) compares the wake function extracted from the CST-calculated wake potential through deconvolution with the analytical wake function. For a *rms* bunch length of $1\,mm$, the extracted wake function shows excellent agreement with the analytical solution in the long-range region, including the oscillatory structure of the wakefield. In contrast, noticeable discrepancies are observed in the short-range region, particularly for $s < 0.4\,mm$. These discrepancies are attributed to numerical limitations in the CST wake-potential calculation. In the short-range regime, the witness particle used for wake-potential evaluation lies within the finite drive-bunch distribution and therefore overlaps spatiotemporally with the wakefields excited by the drive bunch, as illustrated in Fig. 5. As a result, the solver must simultaneously resolve the self-consistent fields of the drive bunch and evaluate the local wake potential at the witness position, which can introduce numerical errors. To examine this effect, simulations with shorter bunch lengths are performed, reducing the spatiotemporal overlap between the witness particle and the fields associated with the drive bunch.

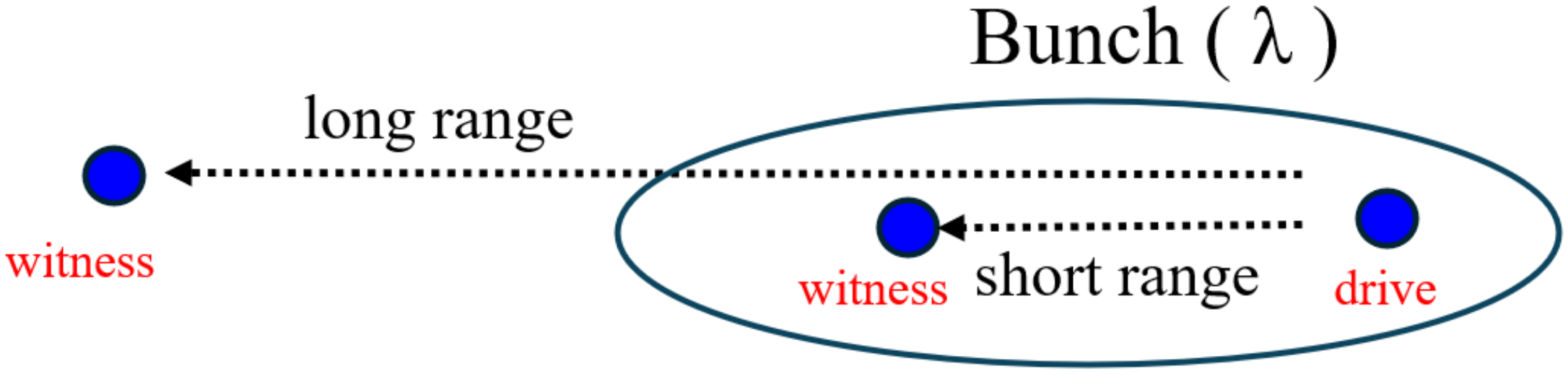

**FIGURE 5.** Schematic illustration of short- and long-range wake potentials excited by a finite bunch. The drive particle is located at the bunch head. In the short-range regime, the witness particle lies within the bunch distribution, whereas in the long-range regime, it is located behind the bunch.

Figure 6 shows the wake functions extracted for the rectangular DLW using different *rms* bunch lengths. As the bunch length is reduced from 0.8 $mm$ to 0.3 $mm$, the extracted short-range wake function approaches the analytical result more closely. The corresponding peak value increases from approximately 2.32 $MV/nC$ to 2.83 $MV/nC$. This trend supports the interpretation that a shorter drive bunch reduces the short-range discrepancy caused by the spatiotemporal overlap between the finite drive-bunch fields and the witness-particle position in the CST wake-potential calculation.

In the region $1\,mm < s < 2\,mm$, increasingly pronounced oscillatory features are observed as the bunch length decreases. This behavior is attributed to the stronger excitation of higher-order modes by shorter bunches. This behavior is consistent with the analytical wakefield expression in Eq. (15), in which the wake function is represented as a superposition of eigenmode contributions. As the bunch length decreases, higher-frequency modal components contribute more significantly, and their superposition produces the observed oscillatory structure. For sufficiently short bunch lengths, these oscillations converge toward the analytical solution, as shown in Fig. 6(d). This result is also consistent with the convolution relation in Eq. (4): as the Gaussian bunch distribution approaches a delta function, the finite-bunch wake potential approaches the point-charge wake function. Accordingly, shorter bunch lengths lead to extracted wake functions that agree more closely with the analytical wake functions.

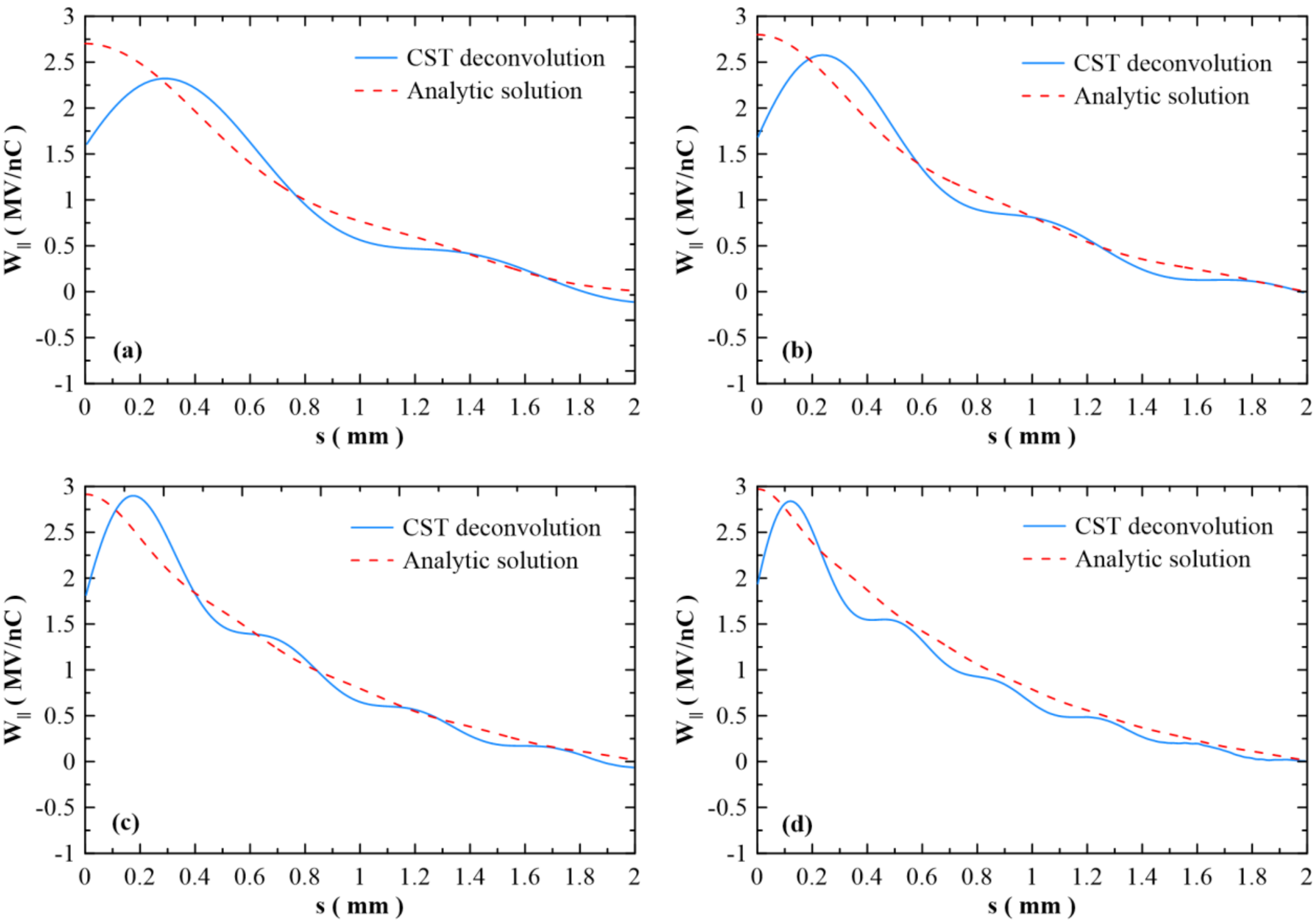

**FIGURE 6.** Comparison between the analytical wake function and the CST-deconvolved wake functions for the rectangular DLW with different rms bunch lengths: (a) $\sigma_\lambda$ = 0.8 *mm*, (b) $\sigma_\lambda$ = 0.6 *mm*, (c) $\sigma_\lambda$ = 0.4 *mm* and (d) $\sigma_\lambda$ = 0.3 *mm*.

The preceding analysis shows that shorter drive bunches improve the recovery of the analytical wake function from CST-calculated wake potentials. However, reducing the bunch length requires finer mesh resolution and substantially increases the computational cost. It is therefore important to evaluate how the reconstructed wake function converges with respect to the bunch length and to identify a practical compromise between accuracy and numerical cost.

In modern accelerator facilities, typical bunch lengths are on the order of 50 $\mu$m to 100 $\mu$m, corresponding to an extremely short temporal scale. As a reference case, we first consider the short-range wakefield excited by an electron bunch with $\sigma_\lambda = 50\ \mu$m in a rectangular waveguide. As listed in Table I, this bunch length corresponds to an excitation bandwidth of approximately 3.546 THz. Figure 7(a) shows that 1197 eigenmodes lie below 10 THz, among which 615 modes are below the 3.546 THz. The same figure also indicates that the lower-order modes have larger amplitudes and therefore make the dominant contribution to the wake function. To quantify the contribution of different modal components, Fig. 7(b) presents the cumulative wake-function amplitude obtained by summing all eigenmode contributions below selected cutoff frequencies. The wake amplitude gradually approaches saturation as more modes are included. For $\sigma_\lambda = 50\ \mu$m, summing the first 615 modes gives a wake-function amplitude of 3.45 MV/nC, but resolving this case in CST would require approximately $3.35 \times 10^{11}$ mesh cells. In comparison, a bunch with $\sigma_\lambda = 100\ \mu$m excites only the first 309 modes and yields a wake-function amplitude of 3.41 MV/nC, corresponding to 98.8% of the $\sigma_\lambda = 50\ \mu$m benchmark case, while reducing the estimated mesh requirement to $4.2 \times 10^{10}$ cells. A further increase of the bunch length to $\sigma_\lambda = 300\ \mu$m reduces the estimated mesh requirement to $2.3 \times 10^{9}$ cells, while still retaining approximately (93.9%) of the benchmark wake amplitude. These results indicate that moderately longer bunches can provide sufficiently accurate wake-function reconstruction with a substantially reduced computational cost. This level of accuracy is also consistent with the IMPACT-T simulations, which show that approximately 90% agreement with the analytical wake function is sufficient to reproduce the beam phase-space evolution.

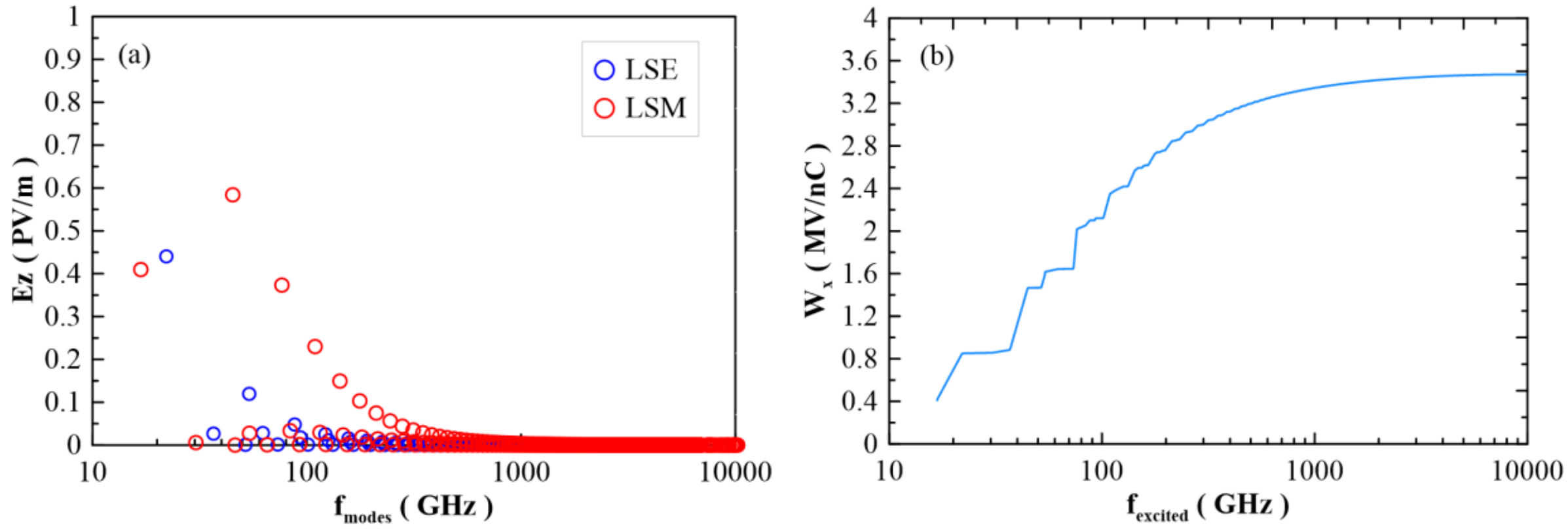

**FIGURE 7.** Modal analysis of the short-range wake function in the rectangular DLW. (a) Electric-field amplitude of each resonant mode. (b) Cumulative wake-function amplitude obtained by summing the contributions of resonant modes below selected cutoff frequencies.

As discussed above, wakefields can be used to design energy dechirpers that reduce the correlated energy spread of electron bunches, thereby improving beam quality for free-electron laser applications. Dielectric-lined waveguide (DLW) dechirpers are particularly attractive because of their compact size, cost effectiveness, and relative ease of fabrication. In this work, a planar DLW dechirper is used as a benchmark case to verify the wake function extracted through the CST-based deconvolution procedure. The extracted wake function is then imported into IMPACT-T for particle-tracking simulations, with a *rms* bunch length of $0.3\ mm$ used as the representative case. The detailed structural parameters of the dechirper are listed in Table III.

IMPACT-T also provides a built-in DLW module that calculates the beam phase-space distribution using the analytical wakefield expressions in Eqs. (13) and (14). The results obtained using the imported wake function are compared with those from the built-in DLW module to verify the consistency and applicability of the proposed method.

TABLE III. Optimized parameters of the flat DLW dechirper.

| Parameter | Value |
| --- | --- |
| a [mm] | 0.5 |
| b [mm] | 3 |
| W [mm] | 10 |
| Dielectric Constant | 4 |
| Total length [mm] | 60 |

As shown in Fig. 8(a), the longitudinal phase-space distributions demonstrate that the energy chirp is effectively suppressed after the beam passes through the DLW dechirper. The initial chirp of approximately $42\ \mathrm{keV}/\mu\mathrm{m}$ is significantly reduced, indicating effective dechirping. Figure 8(b) compares the results obtained using the CST-deconvolved wake function imported into IMPACT-T with those obtained using the built-in DLW module in IMPACT-T. The two results show excellent agreement, confirming the feasibility and consistency of the proposed method.

In the present workflow, the CST wakefield simulation requires only a few hours using approximately $10^9$ mesh cells and 500 GB of RAM, while the subsequent IMPACT-T particle-tracking simulation is completed within a few seconds. This combined CST–IMPACT-T approach therefore provides an efficient method for evaluating DLW dechirper performance while substantially reducing the overall computational cost.

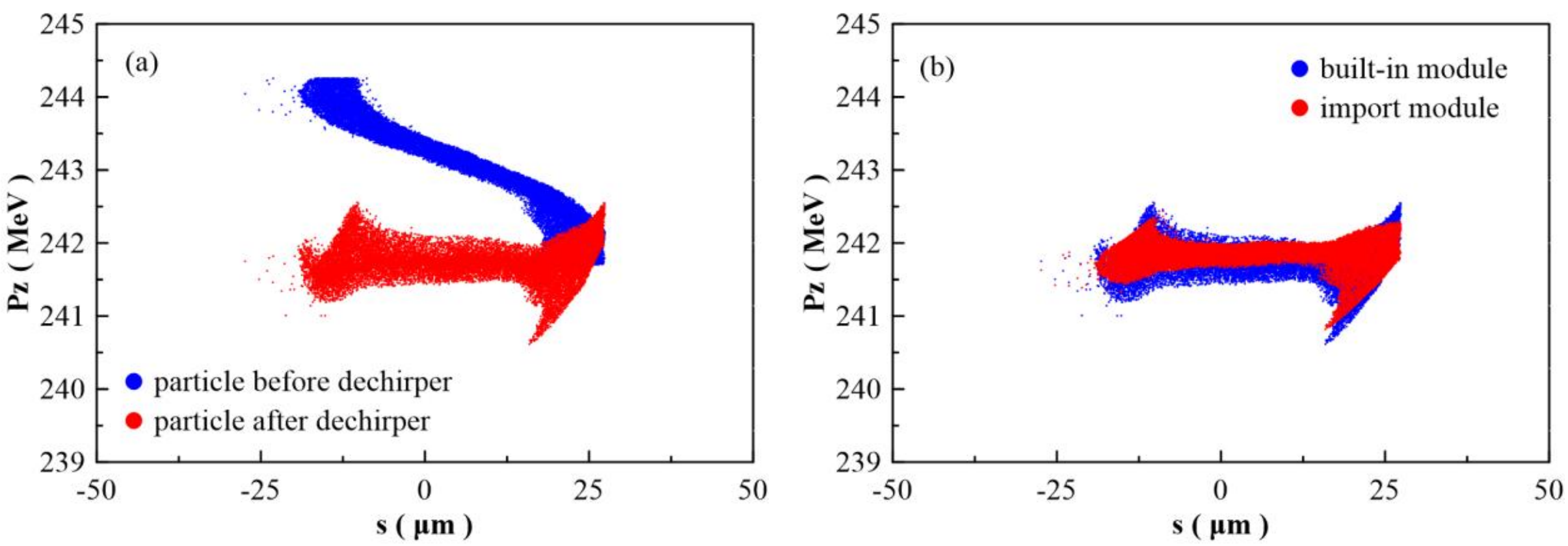

**FIGURE 8.** (a) Longitudinal phase-space distributions before and after the DLW dechirper. (b) Comparison between the built-in IMPACT-T DLW wake-function module and the imported CST-deconvolved wake function.

## V. IMPLICATIONS OF WAKE-FUNCTION EXTRACTION FOR STUDIES OF TRANSVERSE WAKEFIELD MITIGATION

Previous studies on rectangular DLW structures have shown that even a small transverse beam offset can excite substantial transverse wakefields. Analytical solutions further predict focusing and defocusing effects along the two transverse axes [35]. These transverse wakefields may act on trailing particles within the bunch and, in severe cases, drive BBU instabilities that degrade beam quality or lead to beam loss. Earlier approaches for suppressing transverse wakefields relied on two orthogonally oriented planar DLW structures to provide mutual compensation. However, such configurations increase the overall system size and cost and introduce additional challenges in alignment and mechanical stabilization. With the methodology developed in this work, complex geometrical and material distributions can be systematically optimized to reduce transverse wakefields without significantly degrading the longitudinal wakefield.

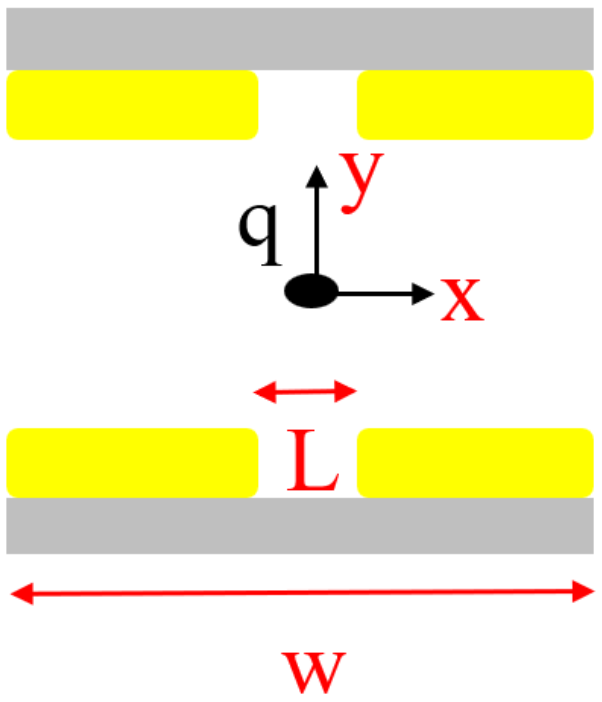

**FIGURE 9.** Layout of the rectangular DLW with a discontinuous dielectric layer. A portion of the dielectric layer is replaced by a vacuum gap with width $L$.

TABLE IV. Optimized structural and material parameters of the modified flat DLW dechirper.

| Parameter | Values |
|---|---|
| a [mm] | 0.5 |
| b [mm] | 3 |
| W [mm] | 10 |
| L [mm] | 0.5 |
| Dielectric Constant | 4 |

Based on this approach, we propose a modified DLW structure with a discontinuous dielectric distribution, as shown in Fig. 9. The corresponding structural parameters are listed in Table IV. In this design, a 0.5 mm -wide section of the dielectric layer near the central region of the DLW cavity is removed and replaced by vacuum. The length L of this modified section is optimized using CST. After optimization, the short-range longitudinal wake function remains nearly unchanged compared with that of the conventional continuous-dielectric structure, as shown in Fig. 10. In contrast, the transverse wakefields are substantially suppressed, as shown in Figs. 11(a) and 11(b). In the $y$ -direction, the maximum short-range transverse wake function is reduced by approximately a factor of two compared with the original design. The transverse wake function in the $x$ -direction shows only a slight increase, while remaining relatively small. These results indicate that the proposed discontinuous dielectric configuration can effectively mitigate transverse wakefield effects and reduce the risk of BBU.

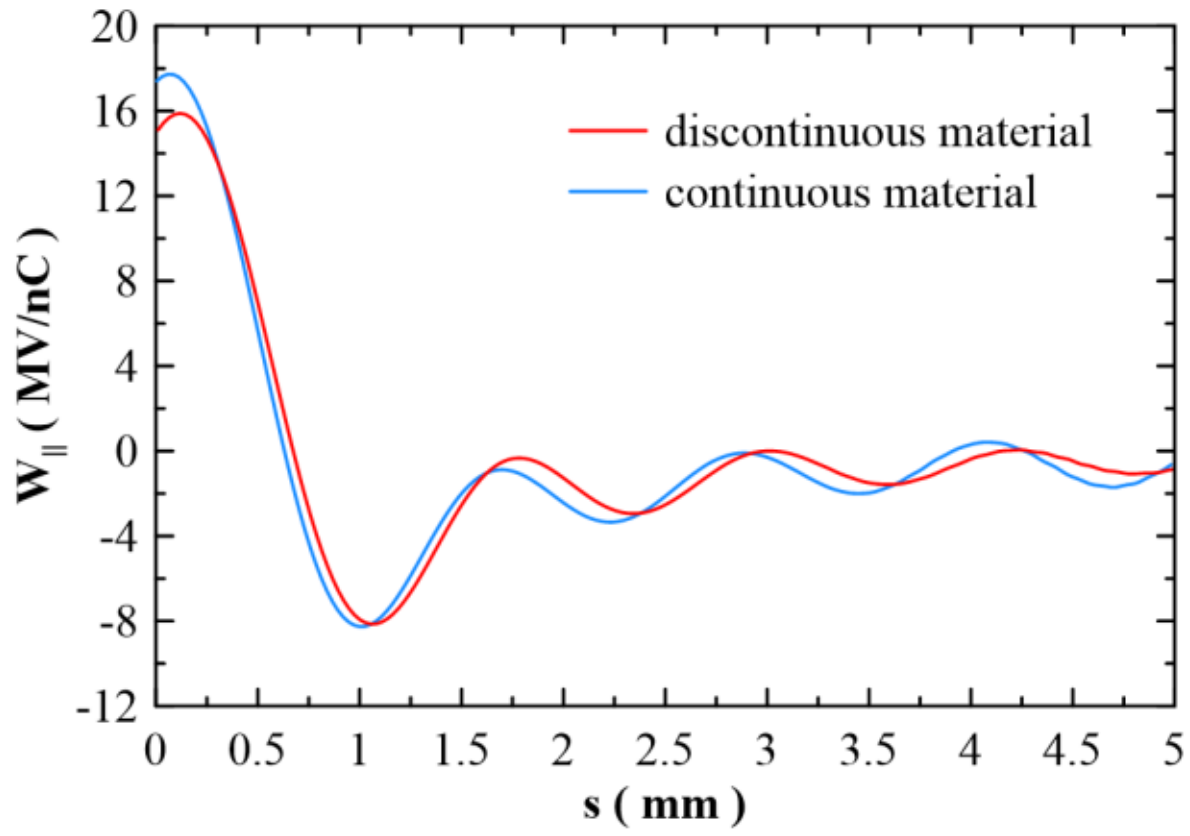

**FIGURE 10.** Comparison of the longitudinal wake functions for DLW structures with continuous and discontinuous dielectric layers, showing that the discontinuous configuration preserves the longitudinal wakefield gradient.

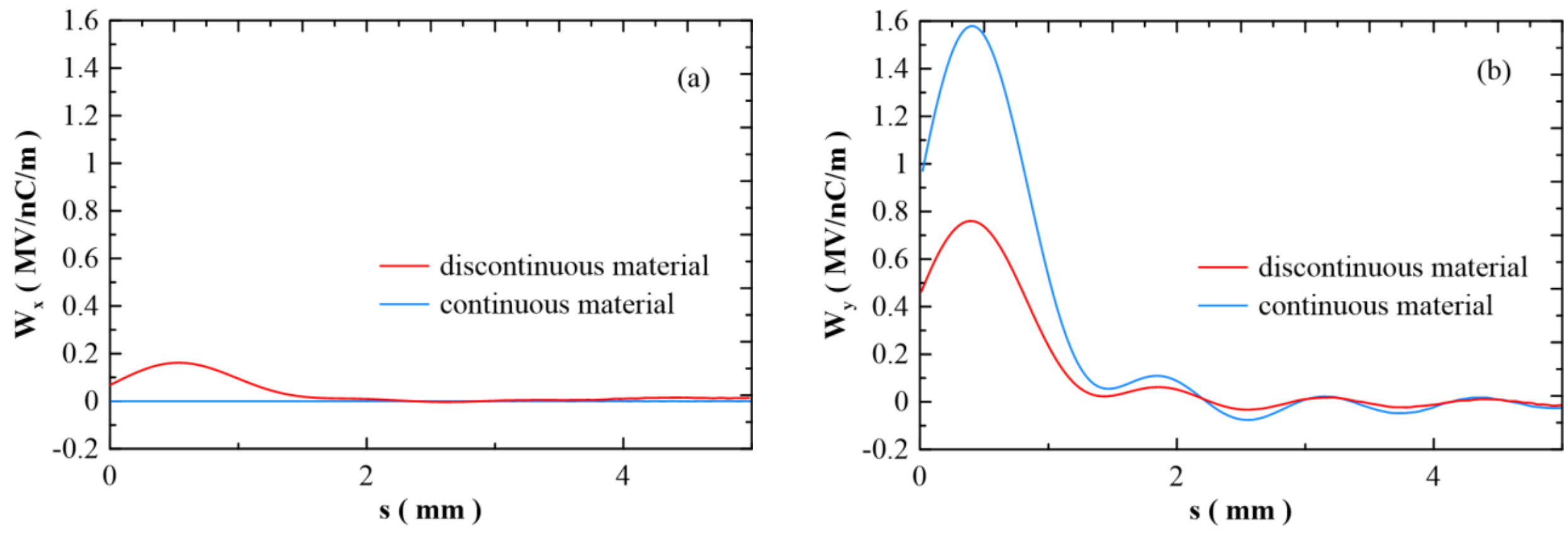

**FIGURE 11.** Transverse wake functions of the continuous- and discontinuous-dielectric DLW structures for particle offsets along (a) the $x$-axis and (b) the $y$-axis.

The corresponding transverse impedances are shown in Figs. 12(a) and 12(b). In the y -direction, the transverse impedance is reduced by approximately one order of magnitude. In the x -direction, the transverse impedance is intrinsically smaller than that in the y -direction because no dielectric layer is present along this direction. Importantly, this discontinuous dielectric configuration cannot be readily treated using analytical models, which highlights the necessity and usefulness of the CST-based wake-function extraction method developed in this work.

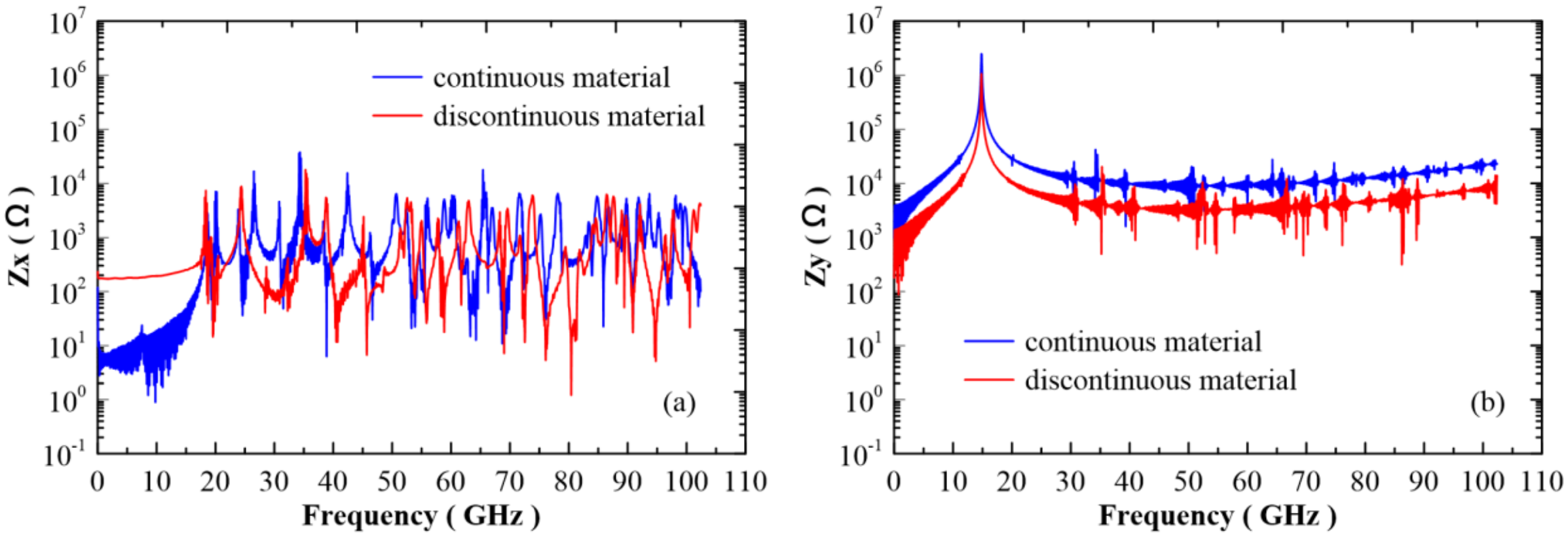

**FIGURE 12.** Transverse impedance with and without structure optimization for particle offsets along (a) the $x$-axis and (b) the $y$-axis.

## VI. CONCLUSION

This study proposes a systematic methodology for extracting wake functions of complex electromagnetic structures from numerically calculated wake potentials. Since mesh-based simulations cannot directly represent a point-charge excitation, the wake potential was first calculated using a finite Gaussian drive bunch in CST, and the corresponding point-charge wake function was then reconstructed through deconvolution with the prescribed bunch distribution. The accuracy of the extracted wake function was shown to depend strongly on the convergence and quality of the calculated wake potential.

The proposed method was first validated using rectangular dielectric-lined waveguide (DLW) structures for which analytical wake functions are available. The wake functions extracted from CST-calculated wake potentials showed good agreement with analytical solutions in both the short- and long-range regimes, provided that a sufficiently short drive bunch was used. The influence of bunch length was examined in detail, showing that shorter bunches improve the recovery of the short-range wake function but require substantially higher mesh resolution and computational resources. This analysis provides a practical criterion for balancing numerical accuracy and computational cost.

The extracted wake function was further implemented in IMPACT-T for multi-particle beam-dynamics simulations of a flat DLW dechirper. The resulting longitudinal phase-space distributions were compared with those obtained using the built-in analytical DLW wake-function module in IMPACT-T. The good agreement between the two approaches confirms that the CST-deconvolved wake function can be reliably used in particle-tracking simulations. This workflow separates the expensive electromagnetic wakefield calculation from the subsequent beam-dynamics calculation, allowing repeated phase-space evaluations to be performed with significantly reduced computational cost.

The method was also applied to the optimization of transverse wakefield suppression in a modified rectangular DLW structure. Motivated by the fact that small transverse beam offsets can excite substantial transverse wakefields and potentially lead to beam-breakup instabilities, a discontinuous dielectric configuration was proposed. In this design, a section of the dielectric layer was removed and replaced by vacuum. The optimized structure preserved a longitudinal wakefield comparable to that of the original continuous-dielectric structure while significantly reducing the transverse wakefields and the corresponding transverse impedances. In particular, the transverse impedance in the $y$ -direction was strongly suppressed, demonstrating the effectiveness of the proposed structural modification.

Because such discontinuous dielectric configurations cannot be readily described using existing analytical models, these results highlight the usefulness of the present CST-based deconvolution method for complex wakefield structures. The proposed approach provides a practical and computationally efficient route for obtaining wake functions, evaluating beam-dynamics effects, and optimizing wakefield devices. Future work will extend this method to slice-energy-spread diagnostics in accelerator facilities and to the design of more advanced wakefield structures for improved beam stability and reduced computational cost.

## ACKNOWLEDGEMENTS

The authors are grateful to Dr. Ji Qiang of Lawrence Berkeley National Laboratory (LBNL) for his useful suggestions regarding the use of the IMPACT-T code in this study. We also thank Shan-You Teng for helpful discussions. This work was partially supported by the National Science and Technology Council (NSTC), Taiwan, under Grant Nos. NSTC 113-2112-M-213-024, NSTC 113-2119-M-001-007, NSTC 113-2112-M-008-010, NSTC 114-2112-M-008 -023 -MY2 and NSTC 114-2112-M-213-009